\documentclass[prb,aps,preprint,showpacs]{revtex4}

\usepackage{epsfig}
\usepackage{natbib}
\usepackage{stmaryrd}

\begin{document}

\title{Spin frustration and magnetic ordering in triangular lattice antiferromagnet Ca$_3$CoNb$_2$O$_9$}

\author{J. Dai$^{1}$}
\author{P. Zhou$^{1}$}
\author{P. S. Wang$^{1}$}
\author{F. Pang$^{1}$}
\author{T J. Munsie$^{2}$}
\author{G. M. Luke$^{2,3}$}
\author{J. S. Zhang$^{4}$}
\author{Weiqiang Yu$^{1}$}
\email{wqyu_phy@ruc.edu.cn}
\affiliation{$^{1}$Department of Physics, Renmin University of China, Beijing 100872, China\\
$^2$Department of Physics and Astronomy, McMaster University, Hamilton L8S 4M1, Canada\\
$^3$Canadian Institute for Advanced Research, Toronto M5G 1Z8, Canada\\
$^{4}$School of Energy, Power and Mechanical Engineering, North China Electric Power University, Beijing 102206, China}

\date{\today}

\pacs{75.10.-Jm, 75.30.-m, 75.30.Cr}

\begin{abstract}

 We synthesized a quasi-two-dimensional distorted triangular lattice antiferromagnet Ca$_3$CoNb$_2$O$_9$, in which the effective spin of Co$^{2+}$ is 1/2 at low temperatures, whose magnetic properties were studied  by dc susceptibility and magnetization techniques. The x-ray diffraction confirms the quality of our powder samples. The large Weiss constant $\theta_{CW}\sim$ $-55$ K and the low Neel temperature($T_N\sim$ 1.45 K) give a frustration factor $f$ ($=\mid\theta_{CW}/T_N \mid$) $\approx$ 38, suggesting that Ca$_3$CoNb$_2$O$_9$ resides in strong frustration regime. Slightly below $T_N$, deviation between the susceptibility data under zero-field cooling (ZFC) and field cooling (FC) is observed. A new magnetic state with 1/3 of the saturate magnetization $M_s$ is suggested in the magnetization curve at 0.46 K. Our study indicates that Ca$_3$CoNb$_2$O$_9$ is an interesting material to investigate magnetism in triangular lattice antiferromagnets with weak anisotropy.

\end{abstract}

\maketitle

\section{Introduction}

Geometrically frustrated classical/quantum magnets, in which the simultaneous minimization of local interaction energies cannot compromise with lattice geometry, have attracted a lot research interest in condensed matter physics due to their novel magnetic orders and exotic excitations\cite{1,2,3}. The two dimension (2D) triangular-lattice Heisenberg antiferromagnet (TLHAF) is a paradigmatic example of frustrated magnet. Theoretical studies have reached a consensus that the ground state of the regular TLHAF with S=1/2 selects the 120$^\circ$ non-collinear structure in the absence of field, which is similar to classical Heisenberg model\cite{4,5,6}. While under external field, the quantum fluctuations can stabilize a novel up-up-down quantum state, with a corresponding 1/3 plateau in the magnetization process\cite{7,8,9,10,11}. This model has given an excellent description on the magnetism of compounds such as Ba$_3$CoSb$_2$O$_9$ \cite{12}, Ba$_3$NiNb$_2$O$_9$\cite{13} and Ba$_3$CoNb$_2$O$_9$ \cite{14}.

For the TLHAF with spatially anisotropic exchange, for example Cs$_2$CuBr$_4$, in addition to the up-up-down phase, many other phases with $\frac{1}{2}, \frac{5}{9}, \frac{2}{3}$ of $M_s$ were observed in the magnetization curve\cite{15}. Theoretical studies suggested that the interplay of spatially anisotropic interactions and quantum fluctuations should play an important role in new phases of Cs$_2$CuBr$_4$\cite{11,15,16}. Therefore, it's of great interest to explore novel magnetic states in distorted triangular lattice antiferromagnets (DTLAF).

\begin{figure*}
\includegraphics[width=13cm]{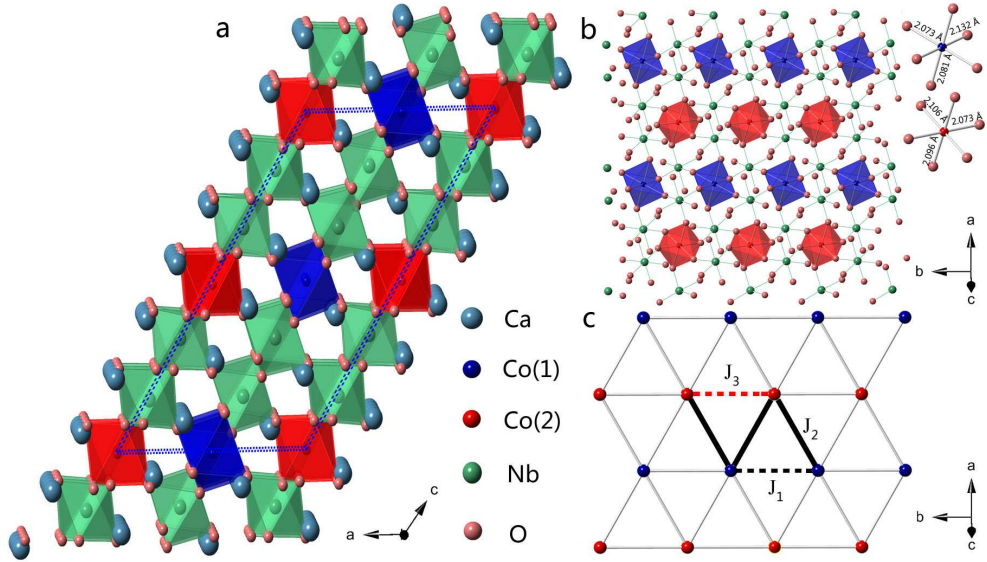}
\caption{\label{structure}(color online) (a) A schematic crystal structure for Ca$_3$CoNb$_2$O$_9$. (b) Two inequivalent Co sites, Co(1) coordinated in blue octahedra environment and Co(2) located in red octahedra, forming the magnetic layers. Local parameters of Co(1)O$_6$ and Co(2)O$_6$ octahedra are also presented in an expanded view. (c) The triangular lattice of Co$^{2+}$ parallel to the ab plane. Three exchange interactions are illustrated: J$_1$, J$_2$ and J$_3$. }
 \end{figure*}

Recently, a new DTLAF material Ca$_3$CoNb$_2$O$_9$ was reported, whose structure was determined by neutron powder diffraction\cite{17}, but its magnetism has not been studied yet. As shown in Fig.~\ref{structure} (a), Ca$_3$CoNb$_2$O$_9$ crystallizes in monoclinic structure (space group $P2_1/c$). There are two inequivalent Co$^{2+}$ ions coordinated in two kinds of twisted CoO$_6$ octahedra which are shown by an enlarged view [see Fig.~\ref{structure} b]. The Co$^{2+}$ ions form almost perfect equilateral-triangular lattice layers parallel to the ab plane and are separated by two nonmagnetic NbO$_6$ octahedra and Ca$^{2+}$ ions. The distance between Co(1)-Co(1),  Co(1)-Co(2), Co(2)-Co(2) are 0.5462 \AA, 0.5521\AA \space and 0.5462 \AA, respectively. Considering the nonuniformity of Co$^{2+}$ sites, there should be three different dominant exchange interactions between nearest neighboring Co$^{2+}$ ions for Ca$_3$CoNb$_2$O$_9$, which are labeled as J$_1$, J$_2$ and J$_3$ respectively [Fig.~\ref{structure} (c)], the values of them need to be determined by future work. Meanwhile, due to the loss of an inversion center between Co(1) and Co(2), Dzyaloshinsky-Moriya (DM) interactions should also present in Ca$_3$CoNb$_2$O$_9$. Therefore, this represents a prototype TLAFM system with spatial anisotropy and DM interactions. In this work, we report our magnetization studies on Ca$_3$CoNb$_2$O$_9$. We find strong frustration and the existence of long-range order antiferromagnetism(LROAFM) at very low temperatures in this system.

\section{Materials and Methods}

We synthesized polycrystalline samples of Ca$_3$CoNb$_2$O$_9$ by a standard solid state reaction method. CaCO$_3$ (5N Alfa), CoO ($>$3N Alfa) and Nb$_2$O$_5$ (4N Alfa) powders with stoichiometric ratio were ground sufficiently and pressed into pellets. After sintering in air at 850 $^\circ$C for 10 hours, the pellets were then calcined at 1200 $^\circ$C for 72 hours with intermediate grinding and re-pelleting. The room temperature powder X-ray diffraction measurement was performed on a Bruker XRD diffractometer with CuK$\alpha$ radiation. The
  susceptibility measurements in the temperature range 2 K - 300 K were performed with vibrating sample magnetometer (VSM) loaded on a physical property measurement system (Quantum Design PPMS). Magnetic measurements below 2 K were performed using a magnetic property measurement system (Quantum Design MPMS) with a $^3$He insert at McMaster University.

\section{Results}

\begin{figure}
\includegraphics[width=10cm]{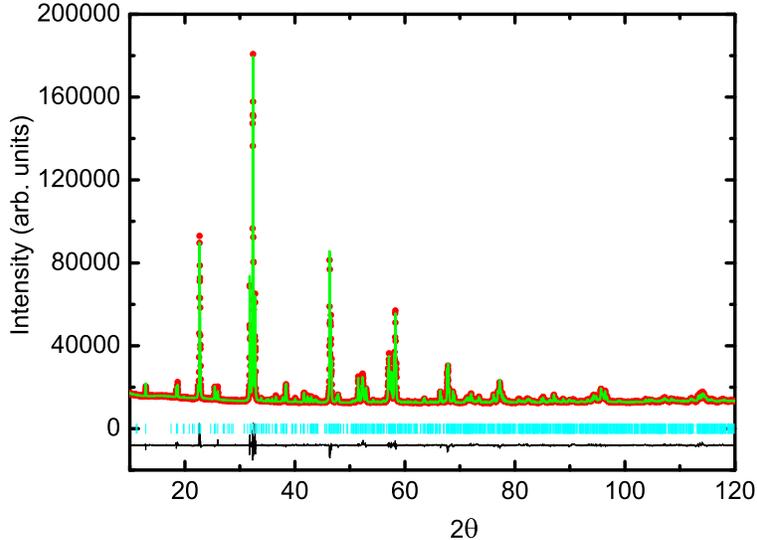}
\caption{\label{XRD}(color online) The room-temperature XRD pattern of polycrystalline Ca$_3$CoNb$_2$O$_9$ samples. The red solid cycles are our experimental data. The green solid line is the calculated XRD pattern for powder Ca$_3$CoNb$_2$O$_9$. The vertical lines indicate the reflection positions for Ca$_3$CoNb$_2$O$_9$. The difference curve is also presented as the black solid line. }
\end{figure}

   The XRD patterns of polycrystalline Ca$_3$CoNb$_2$O$_9$ samples are presented in Fig.~\ref{XRD}. Refinement of the observed diffraction peaks using the structure parameters obtained from previous neutron diffraction\cite{17} gives a final weighted residual error Rwp = 2.64, which means that our XRD data match well with the calculated results. The obtained structure parameters a = 9.5988(9) \AA, b = 5.4576(3) \AA  \space  and  c = 16.8643(9) \AA \space  are also consistent with previous results\cite{17}. No additional impurity peak is found in the whole diffraction range from 5$^\circ$ to 120$^\circ$ indicating that there is no obvious impurity phase in our samples.

\begin{figure}
\includegraphics[width=10cm]{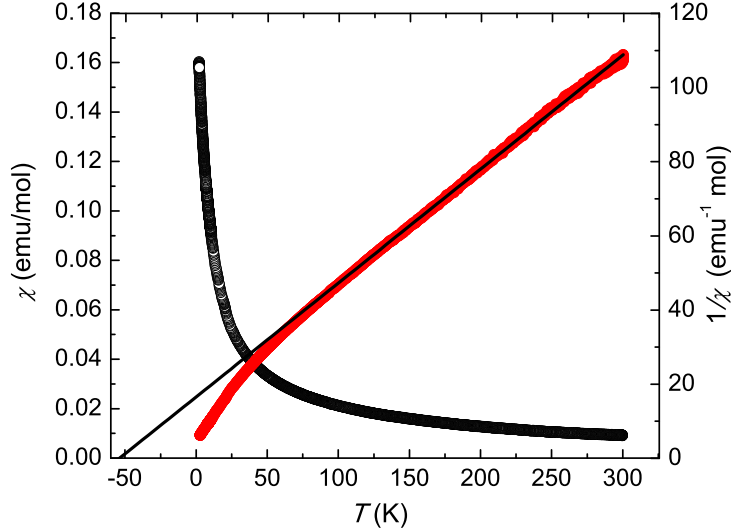}
\caption{\label{sus}(color online) The dc susceptibility ($\chi= M/H$) and its inverse measured at 0.1 T. The solid line is a linear Curie-Weiss fitting of $\chi$ in temperature range 100 -300 K.}
\end{figure}

The dc susceptibility and its inverse above 2 K under a field of 0.1 T for Ca$_3$CoNb$_2$O$_9$ are shown in Fig.~\ref{sus}. The 1/$\chi$ deviates from the Curie-Weiss curve at around 50 K , which is because of the frustration when T$< \theta_{CW}$ . The susceptibility data above 100 K can be fit by the Curie-Weiss formula: $\chi_M$ = C/($T-\theta_{CW}$)+$\chi_0$, where C is the Curie constant,   $\theta_{CW}$ is the Weiss constant, and the $\chi_0$ is the temperature-independent paramagnetic susceptibility. The fitting gives an effective moment of 5.1 $\mu_B$ and a Weiss constant of $\theta_{CW}\sim$ $-55$ K. Considering the contribution from the orbital angular momentum, the value of the moment is reasonable for the Co$^{2+}$ ion in the high-spin state\cite{18}. The negative Weiss constant reflects dominant antiferromagnetic interactions between Co$^{2+}$ ions. The large Weiss temperature and the absence of magnetic ordering down to 2 K imply the presence of strong frustration in this system.

\begin{figure}
\includegraphics[width=13cm]{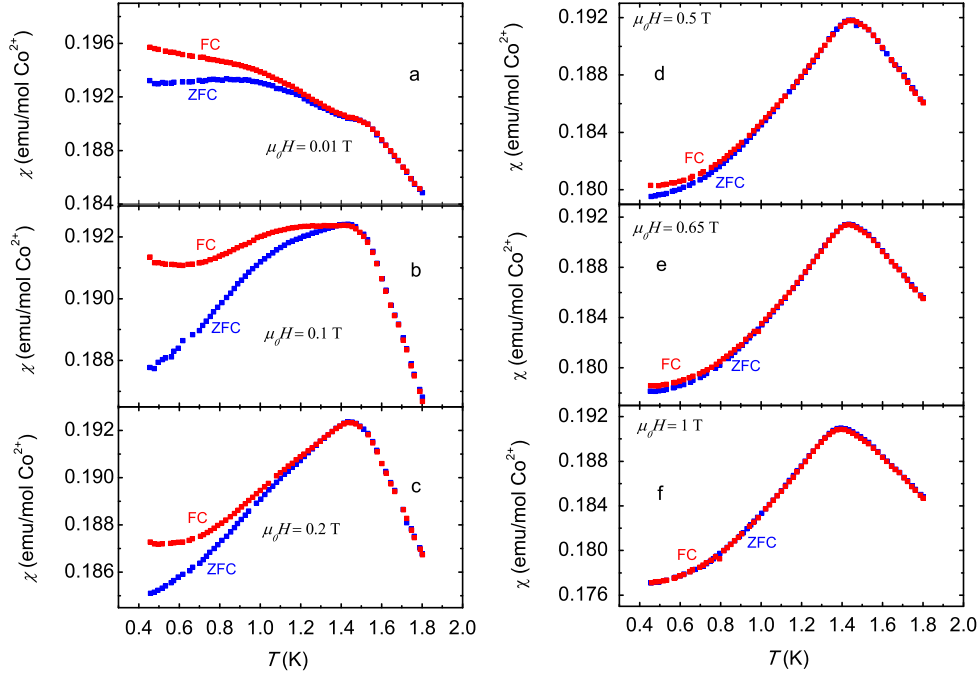}
\caption{\label{SUSvsH3}(color online) (a)-(f) dc susceptibility down to 0.46 K after ZFC and FC measured at 0.01 T,  0.1 T,  0.2 T,  0.5 T,  0.65 T  and  1 T respectively. }
 \end{figure}

To study the ground state magnetism of Ca$_3$CoNb$_2$O$_9$, we performed susceptibility measurements down to 0.46 K with ZFC and FC methods. As shown in Fig.~\ref{SUSvsH3} (a), under a 0.01 T external field,  both ZFC and FC susceptibility increase with cooling. With further cooling, a cusp occurs at around 1.45 K, which indicates a transition from the paramagnetic phase to an ordered state. The susceptibility is less than 0.2 emu/mol Co$^{2+}$ in the ordered state, which suggests an long-range order antiferromagnetism (LROAFM). Combined with the large Weiss constant obtained above, the frustration factor $f$(=$\mid\theta_{CW}\mid/T_N$) is about 38, which is comparable to that of strong frustrated regular TLHAF Ba$_3$CoNb$_2$O$_9$ ($\theta_{CW}$= $-51$ K and $T_{N1}$ = 1.36 K)\cite{14}. In contrast to the regular triangular lattice antiferromagnet Ba$_3$CoNb$_2$O$_9$, Ca$_3$CoNb$_2$O$_9$ has DM interactions and spatially anisotropic exchange interactions. Theoretical studies suggested that both interactions release frustration and enhance the LROAFM in general\cite{19,20}. For Ca$_3$CoNb$_2$O$_9$, however, the anisotropy does not lift up magnetic ordering temperature significantly, which suggests that the DM interactions and spatial anisotropy are weak.

The magnetically ordered ground states were expected for the insulating triangular lattice antiferromagnets, including isotropic Heisenberg\cite{14}, XXZ$^{[21]}$ and distorted \cite{11}models. Experimentally, almost all triangular antiferromagnets have been found with LROAFM at low temperatures, except $\kappa$-(BEDT-TTF)$_2$Cu$_2$(CN)$_3$, which is probably closer to a Hubbard model\cite{22}. Therefore, the LROAFM state is natural for our triangular lattice antiferromagnet.

At T $\sim$ 1.2 K, which is slightly below $T_N$, deviation appears between the ZFC and FC data under 0.01 Tesla field [Fig.~\ref{SUSvsH3}. (a)]. We further measure dc susceptibility after FC and ZFC at higher fields[see Fig.~\ref{SUSvsH3} (b)-(f)]. Under a 0.1 Tesla field, the FC susceptibility is suppressed slightly below T$_N$, while the suppression effect is much stronger for the ZFC data, resulting in an enhancement of the deviation between the FC and ZFC data. With further increasing fields, the discrepancy is gradually suppressed  monotonously and completely removed by a field of 1 T, while the position of $T_N$ remains at 1.45 K.

The deviation between ZFC and FC susceptibility data at low measurement fields was reported in some spin-canted antiferromagnets. For example,  in DTLAFM Cu$_2$(OH)$_3$(C$_m$H$_{2m+1}$COO), m = 7, 9, 11\cite{23} the DM interactions were suggested to play an essential role to the spin canting. A similar situation may occur in Ca$_3$CoNb$_2$O$_9$, where the DM interactions assist the canting of antiferromagnetic moments.

 The bifurcation of ZFC and FC susceptibility was also seen in spin glass system with spin-freezing. Earlier theoretical studies suggested that the spin glass phase usually occurs in frustrated systems with quenched disorder\cite{24,25}. However, spin glass can also be realized in system with competing ferro-and antiferromagnetic bonds distributed in a certain rule, rather than a random way\cite{26}. Experimentally, the spin freezing transition has been found recently in several nominally disorder-free frustrated systems such as Y$_2$Mo$_2$O$_7$\cite{27,28}, Gd$_3$Ga$_5$O$_{12}$\cite{29}, ZnCr$_2$O$_4$\cite{30}, CsVCl$_3$\cite{31},  Cu$_2$(OH)$_3$(C$_m$H$_{2m+1}$COO), m = 7, 9, 11\cite{23}. The spin glass and LROAFM phases coexist in the latter three materials. Whether disorder is essential to the generation of spin glass in a strongly frustrated magnet remains to be established. Structurely, Ca$_3$CoNb$_2$O$_9$ is a site ordered system which has been confirmed by neutron diffraction experiment\cite{17}. However, we can not exclude the possibility of mixing tiny amounts of Nb$^{5+}$/Co$^{2+}$ on the perovskite $B$-sites. It has been expected that spin glass can be stabilized in geometrically frustrated antiferromagnets even with smallest degree of disorder\cite{32}. After all, based on our current dc susceptibility data, we cannot exclusively determine the origin of observed deviation between the ZFC and FC susceptibility. Specific heat and ac susceptibility experiments down to $^3$He temperatures may help to resolve this issue.

\begin{figure}
\includegraphics[width=10cm]{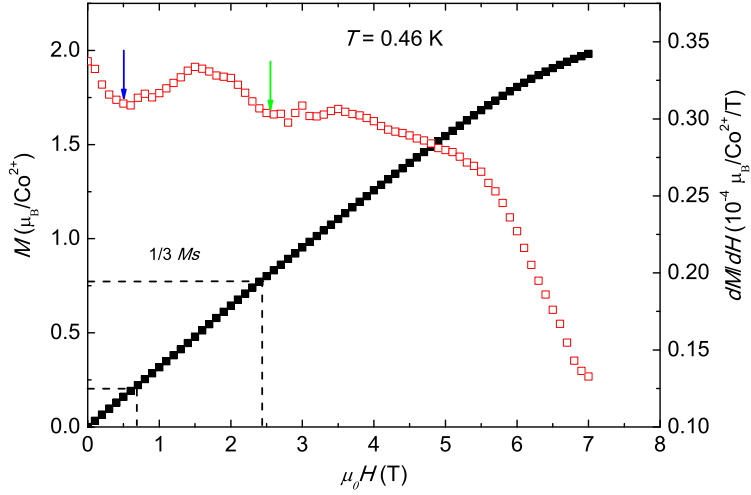}
\caption{\label{MH}(color online) The magnetization curve of Ca$_3$CoNb$_2$O$_9$ at a fixed temperature 0.46 K. The solid squares are M-H curve and the empty ones are the first deviation of magnetization curve as a function of field. The vertical arrows indicate the positions of the valleys. }
 \end{figure}

To investigate the evolution of magnetism with external field in Ca$_3$CoNb$_2$O$_9$, the dc magnetization was measured at 0.46 K [Fig.~\ref{MH}]. The magnetization increases linearly with field and then changes slope around 6 T, which corresponds to a crossover from a long-range ordered state to a fully polarized state, as in Ba$_3$CoNb$_2$O$_9$\cite{14,33}. Above 6 T, though the magnetization still increases, it has a tendency to saturate around 2 $\mu_B$/Co$^{2+}$. Due to spin-orbital couplings and the crystal field, the ground state of Co$^{2+}$ ion coordinated in octahedral environment is a Kramers doublet with an effective spin of 1/2 \cite{34,35}. The low spin state of Co$^{2+}$ has also been observed in many other TLAFM with octahedral Co sites, such as Ba$_3$CoNb$_2$O$_9$\cite{14}, Ba$_3$CoSb$_2$O$_9$\cite{12} and CsCoCl$_3$\cite{2}. We further plot the first derivative of the magnetization curve as a function of field. Below 6 T, two valleys are observed in the derivative curve, which is reproducible on different samples. The first one is located around 0.7 T, which is consistent with the field where the bifurcation between ZFC and FC susceptibility disappears [Fig.~\ref{SUSvsH3} (b)-(f)]. The second one occurs at around 2.5 T with the magnetization of 0.75 $\mu_B$/Co$^{2+}$, which is close to a 1/3 of the saturation magnetization ($Ms$). The possible appearance of a 1/3 magnetization plateau is consistent with the fact of strong quantum fluctuations from frustration and small effective spins in this triangular lattice\cite{7,8}. However, we are aware that the magnetization plateau is not strong, which may be because our samples are randomly aligned powders. Single crystal samples are needed to confirm this observation.

Our study revealed that the ground state for Ca$_3$CoNb$_2$O$_9$ is a LROAFM coexisting with spin canting or spin glassiness. However, for a regular triangular lattice model, the ground state is coplanar 120$^\circ$ antiferromagnetic order, which has been confirmed in Ba$_3$CoSb$_2$O$_9$\cite{12}, Ba$_3$NiNb$_2$O$_9$ \cite{13} and Ba$_3$CoNb$_2$O$_9$\cite{14}. The different ground state of Ca$_3$CoNb$_2$O$_9$ may be a result of the interplay of strong frustration and anisotropic interactions, and provides possibility of searching for new novel quantum states in this geometrically frustrated magnet. Moreover, the 1/3 plateau in magnetization for a DTLAF is a macroscopic manifestation of quantum effect predicted by theoretical studies\cite{11}. To be more unique, this novel magnetic state can be stabilized by a very small field (~2.5 T) which is easily accessible by experiments, and provides a promising material for further investigation of magnetization plateaus which are very interesting theoretically.
\section{Conclusion}

In conclusion, the magnetism of distorted triangular lattice antiferromagnet Ca$_3$CoNb$_2$O$_9$ has been studied for the first time. The large Weiss constant and very low ordering temperature imply strong geometrical frustration in Ca$_3$CoNb$_2$O$_9$.  Slightly below T$_N$, a deviation between susceptibility data after ZFC and FC is observed below 0.7 T, which may be related to DM interactions and/or site disorder. A magnetic state with 1/3 of $M_s$ is suggested in magnetization curve, which may be caused by strong quantum fluctuations in this triangular lattice. Ca$_3$CoNb$_2$O$_9$ supplies a new system to study the magnetism of triangular lattice antiferromagnet with weak spatially  anisotropic interactions and DM interactions.

Work at the RUC is supported by the NSF of China (Grant No. 11374364 and No. 11222433) and by the National Basic Research Program 
of China (Grant No. 2011CBA00112). Research at McMaster University is supported by the Natural Sciences and Engineering Research Council.
Work at North China Electric Power University was supported by the Scientific Research Foundation for the Returned Overseas Chinese Scholars, State Education Ministry.

\end{document}